\documentclass[a4paper]{jpconf}
\usepackage{graphicx}
\usepackage{bm}
\usepackage{amsmath}
\usepackage{xcolor}
\usepackage{citesort}
\usepackage{cite}

\begin{document}
\title{Connecting Inverse Design with Experimentally Relevant Models}

\author{Beth A Lindquist}

\address{Theoretical Division, Los Alamos National Laboratory, Los Alamos, New Mexico 87545, USA}

\ead{bethl@lanl.gov}

\begin{abstract}
While colloids are promising building blocks for the self-assembly of materials with novel microstructures, their numerous tunable parameters inhibit brute force searching for appropriate parameter combinations that yield self-assembly of a desired structure. Instead, inverse approaches that invoke a systematic optimization framework can effectively navigate this design space. In this proceeding, we apply one such inverse technique, Relative Entropy Minimization, to discover isotropic pairwise interaction potentials that prompt self-assembly of clusters \emph{in silico}. The functional form of the pair interaction is chosen to model a mixture of charged colloids and neutral polymers that act as depletants, and the parameters are directly connected to experimentally tunable quantities.  
\end{abstract}

\section{Introduction}
A promising route to fabricate novel materials is to select particles that mutually interact in such a way that they spontaneous assemble into desirable microstructures. Colloids suspended in solvent possess many tunable parameters (shape, size dispersity, salt concentration, etc.) and therefore are excellent candidates for the self-assembly of a variety of complex structures.~\cite{Colloids1,Colloids2,Colloids3} However, the associated large design space can be difficult to navigate in a trial-and-error fashion. As a result, so-called inverse design approaches, which employ some kind of systematic optimization scheme in order to find optimal parameters for a particular design goal, have been highly useful in the discovery of interaction potentials that result in the self-assembly of interesting structural motifs.~\cite{IDSA1,IDSA2,IDSA3,IDSA4,IDSA5,IDSA6,IDSA7,IDSA8,IDSA9}

In particular, approaches that have been adapted from the molecular coarse-graining literature~\cite{IBI_and_RE_ID} have enabled the discovery of potentials that prompt self-assembly of exotic structures \emph{in silico}, such as clusters,~\cite{IBI_clusters,RE_clusters_pores_crystals,RE_clusters_presssure} pores,~\cite{IBI_RE_pores,IBI_pores2,RE_clusters_pores_crystals} crystals,~\cite{RE_crystals,binaryRE,RE_clusters_pores_crystals,RE_FK,RE_crystals_kspace} and strings~\cite{RE_strings}. Relative Entropy (RE) Minimization is one such approach; in this context, the RE quantifies the statistical distance between two ensembles of configurations.~\cite{general_RE,RE_ID,IBI_and_RE_ID} When one set of configurations possesses a targeted structural motif and the other set is sampled from a tunable probabilistic model, then minimization of the RE will maximize the likelihood that the model system will reproduce the targeted microstructures. In molecular coarse-graining, the target ensemble is typically the output of an all-atom simulation, and the model represents multiple atoms as a single bead. For self-assembly, the target ensemble has often been constrained in an unphysical way to possess a desired microstructure, and a (often pairwise isotropic) tunable interaction potential sampled according to Boltzmann statistics is the probabilistic model. 

While RE Minimization has proven a powerful tool to discover interaction potentials relevant to self-assembly, the resulting interactions are often not straightforward to realize experimentally. However, the RE framework (described, along with other computational considerations, in Sect.~\ref{sec:methods}) is easily extensible to interaction forms that have a tangible connection to experiment. In this proceeding, we make progress towards applying RE Minimization to more experimentally relevant systems by optimizing the parameters associated with the Asakura-Oosawa (AO) model for depletion attraction combined with the Derjaguin-Landau-Verwey-Overbeek (DLVO) model for electrostatic repulsion. We optimize the parameters of the AO+DLVO model to discover interaction potentials that drive the self-assembly of clustered configurations, that is, finite aggregates of a particular size. It is well established that interactions possessing a short-ranged attractive well in competition with a longer-ranged repulsive barrier can drive the formation of clusters~\cite{SALR1,SALR2,SALR3,SALR4,SALR5,SALR6,SALR7,SALR8,SALR9,SALR10,SALR11,CSD_1,CSD_2}, and depletion attraction in combination with electrostatic repulsion is one typical route to realize these physics in real systems.~\cite{Real1,Real2,Real3,Real4,Real5} In this proceeding, we show that it is possible to inversely design for clusters that are of approximately the desired size in this parameter space. These results are described in Sect.~\ref{sec:results}, and we conclude in Sect.~\ref{sec:conclusions}. 

\section{Computational Methods} \label{sec:methods}

\subsection{Functional Form of Interaction}

The functional form used in the optimization is a combination of a smoothed effective AO model for depletion attraction and the DLVO model for electrostatic repulsion. The DLVO interaction is given by
\begin{equation} \label{eqn:dlvo} 
\beta u_{\text{DLVO}}(r)=A\frac{e^{-r/z}}{r/\sigma}
\end{equation} 
where $A$ is a dimensionless energy related to the surface charge of the colloids, $z$ is the screening length due to the salt concentration of the solvent, and $\sigma$ is the colloid diameter.

The AO model represents a mixture of colloids and polymers; however, when the polymers are sufficiently small relative to the colloids, the polymer degrees of freedom can be integrated out to arrive at a single-component effective AO model. The smoothed effective AO interaction is the sum of an attractive well and a hard-core-like component. The former is given by 

\begin{equation} \label{eqn:ao}
\beta u_{\text{AO}}(r)=
\begin{array}{ll}
      \eta^r_p (1+q^{-1})^3 \big[1-\frac{3r/ \sigma}{2(1+q)}+\frac{(r/ \sigma)^3}{2(1+q)^3}\big], & \sigma<r<\sigma+\sigma_p \\
      0, & r<\sigma \text{ or } r>\sigma+\sigma_p \\
\end{array} 
\end{equation} 
where $\eta^r_p$ is the reservoir packing fraction of the polymer (that sets the chemical potential) and $q$ is the ratio of the polymer diameter ($\sigma_p$) to $\sigma$. 

The smooth hard-core-like component is described by

\begin{equation} \label{eqn:core} 
\beta u_{\text{core}}(r)=4\Big(\Big[\frac{b \sigma}{r-\epsilon \sigma}\Big]^{12}+\Big[\frac{b \sigma}{r-\epsilon \sigma}\Big]^{6}-\Big[\frac{b \sigma}{b \sigma + q -\epsilon \sigma}\Big]^{12}-\Big[\frac{b \sigma}{b \sigma + q -\epsilon \sigma}\Big]^{6}\Big)
\end{equation} 
where $b=0.01$ and $\epsilon=0.98857$.~\cite{smoothAO} The total potential is derived from the sum of Eqs.~\ref{eqn:dlvo}--~\ref{eqn:core}. Since Eq.~\ref{eqn:core} furnishes a very steep repulsion, any values of the corresponding force that exceed $1000 k_bT/ \sigma$ are set to $1000 k_bT/ \sigma$. The new force is then integrated with respect to $r$ to arrive at the final potential form; the potentials generated via the above procedure are smooth and continuous. The parameters $\eta^r_p$, $q$, $A$, and $z$ comprise the parameters $\boldsymbol{\theta}$ to be optimized via RE Minimization.  

\subsection{Relative Entropy Minimization}

We minimize the RE via an iterative scheme. First, an initial guess potential is simulated. Using both the resulting simulation and the target ensemble, the gradient of the RE with respect to the tunable parameters $\boldsymbol{\theta}$ can be computed to arrive at a new interaction potential that is used as input to the next simulation. This iterative procedure is carried out until a potential fulfilling the design criteria is found. RE Minimization can be performed for arbitrarily complex interaction forms; however, for isotropic pairwise interactions ($u(r|\boldsymbol{\theta})$) simulated in the NVT ensemble in three dimensions, the update scheme for the parameters is particularly simple:~\cite{RE_clusters_pores_crystals}
\begin{equation} \label{eqn:update} 
\boldsymbol{\theta}^{(i+1)}=\boldsymbol{\theta}^{(i)}+\boldsymbol{\delta} \int_{0}^{\infty} dr \frac{r^2}{\sigma^3} \big[g(r| \boldsymbol{\theta}^{(i)})-g_{\text{tgt}}(r)\big] \big[\boldsymbol{\nabla}_{\boldsymbol{\theta}} \beta u(r|\boldsymbol{\theta})\big]_{\boldsymbol{\theta}=\boldsymbol{\theta}^{(i)}} 
\end{equation} 
where $g(r| \boldsymbol{\theta}^{(i)})$ and $g_{\text{tgt}}(r)$ are the radial distribution functions of the potential at step $i$, $u(r|\boldsymbol{\theta}^{(i)})$, and the target ensemble, respectively, $\beta = (k_b T)^{-1}$ where $k_b$ is Boltzmann's constant and $T$ is temperature, and $\boldsymbol{\delta}$ are the empirically determined learning rates that are independently set for each of the four tunable parameters. 

All parameters in the optimization were constrained to be positive. Additionally, $q$ was constrained to be less than or equal to 0.25 to stay within a range where we expect the effective AO model to be reasonably accurate. The effective AO model is only technically exact for $q<0.1547$, though prior work has found excellent agreement between the effective and explicit AO model at $q=0.2$ as well.~\cite{q_AO} The initial guess was $\eta^r_p=0.4$, $q=0.2$, $A=0.1$, $z/\sigma=1.0$ or 2. The learning rates were occasionally manually varied in response to the behavior of the optimization, but the starting values for $\boldsymbol{\delta}$ were of order 0.1 for $z$ and 0.001 the other parameters. Typically, in the beginning of the optimization, the simulations oscillate between a homogeneous fluid and either very large clusters or a phase-separated state before settling into clustered configurations. Once the optimization entered a parameter space consistent with clusters, the learning rates for $A$ and $z$ could be increased by a factor of 5--10. About 50--100 optimization cycles were required before the agreement between the simulation and the target stopped improving.

\subsection{Target Simulation}

\begin{figure}[h]
\includegraphics[width=8cm, keepaspectratio]{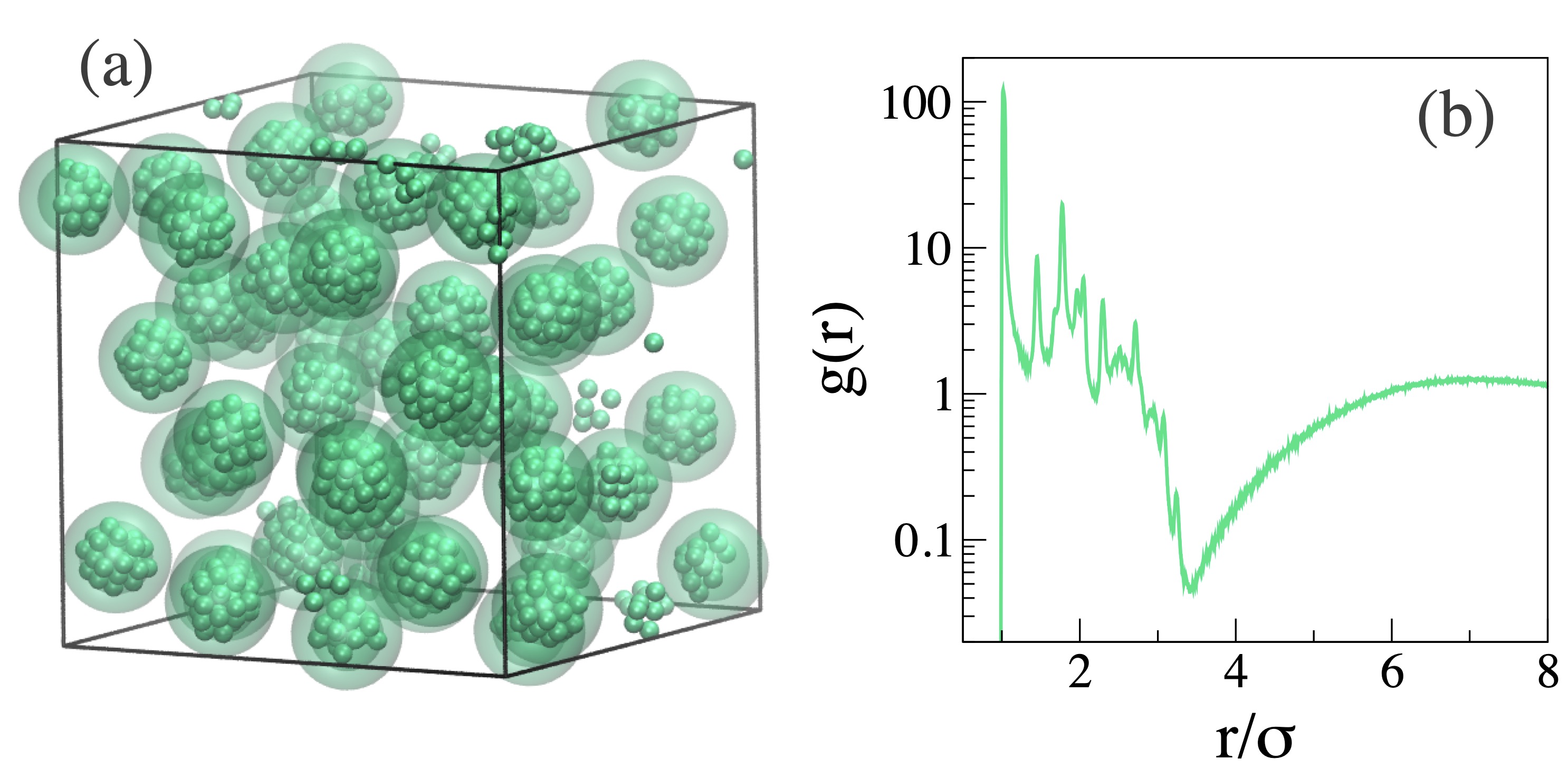}\hspace{0.8cm}%
\begin{minipage}[b]{7cm}\caption{\label{fgr:tgt}a) Representative snapshot from an $n_{\text{tgt}}=32$ target simulation showing the range of the attractive well (inner shell) that constrains the A particles to remain in the cluster and the intercluster repulsion (outer shell) that maintains spatial separation between clusters. (b) Radial distribution function corresponding to the target simulation in panel (a)}
\end{minipage}
\end{figure}

In order to construct an ensemble of configurations where the particles are clustered, we simulate a two-component system where one species (A) can be though of as the ``real'' particles and the second species (B) enforces that the clusters remain as discrete objects. The A particles mutually interact via a hard-core-like WCA term combined with a short-ranged attractive well. The depth of the attractive well ($\epsilon_{\text{att}}^{(A,A)}$) controls the degree of crystallinity of the particles within each cluster. From an initial configuration that contains clusters, one B particle is placed at the center of each cluster that furnishes an attractive well that is sufficiently broad and deep to keep the particles confined to a given cluster. Finally, the B particles interact repulsively via the WCA interaction to ensure that the A particles in different clusters cannot get close enough to interact attractively. A snapshot from the target simulation for $n_{\text{tgt}}=32$ monomers per cluster is shown in Fig.~\ref{fgr:tgt}a, where the smaller AB attractive wells and the larger BB repulsions are represented as translucent spheres and the A particles are opaque. The corresponding $g_{\text{tgt}}(r)$ is shown in Fig.~\ref{fgr:tgt}b, where the pronounced dip at $r\approx 3.5\sigma$ is due to the discrete nature of the clusters. Further details about the target simulations are given in the Appendix.

\subsection{Simulation and Analysis Details}

Simulations were performed in \textsc{gromacs}, version 5.1.2~\cite{GROMACS_1,GROMACS_2} in three dimensions with periodic boundary conditions in all directions. Owing to the steepness of the smoothed effective AO interaction, it was necessary to use a very small timestep of $dt=0.0001 \sqrt{\beta m \sigma^2}$, where $m$ is the particle mass. Simulations were performed in the NVT ensemble, with a canonical velocity re-scaling thermostat~\cite{thermostat} employed to control the temperature where $\tau=250dt$. For all simulations, a packing fraction $\eta=\frac{\pi}{6}\rho \sigma^3$, where $\rho$ is number density, of 0.05 was used. Within the iterative RE cycles, simulations with 512 particles were 20,000,000 steps in length with configuration data saved every 12,500 steps and and statistics collected from the latter half of the configurations. The interaction potential was truncated when the magnitude of the force dropped below $0.002k_bT/\sigma$ For the production runs, the simulations contained 1600 particles, the potential was truncated at the point when the force was less than $0.001k_bT/\sigma$, and the simulation was run for 50,000,000 steps with statistics collected from the last 20\% of the configurations.

The cluster size distribution $p(n)$ gives the probability that an aggregate will be $n$ particles in size.~\cite{CSD_1,CSD_2} A cluster is defined as a group of particles that are all either direct neighbors (here, neighboring particles have an interparticle separation of less than $r_{\text{c}}$) or are connected via a contiguous pathway of neighboring pairs. In general, the appropriate value for $r_{\text{c}}$ depends on the interaction potential; here, we select $r_{\text{c}}=1.08\sigma$ such that it approximately coincides with the point at which the depth of the attractive well is at half of its minimum value. 

\section{Results and Discussion} \label{sec:results}

\subsection{Control of cluster size}

\begin{figure}[!htb]
  \includegraphics[width=16cm, keepaspectratio]{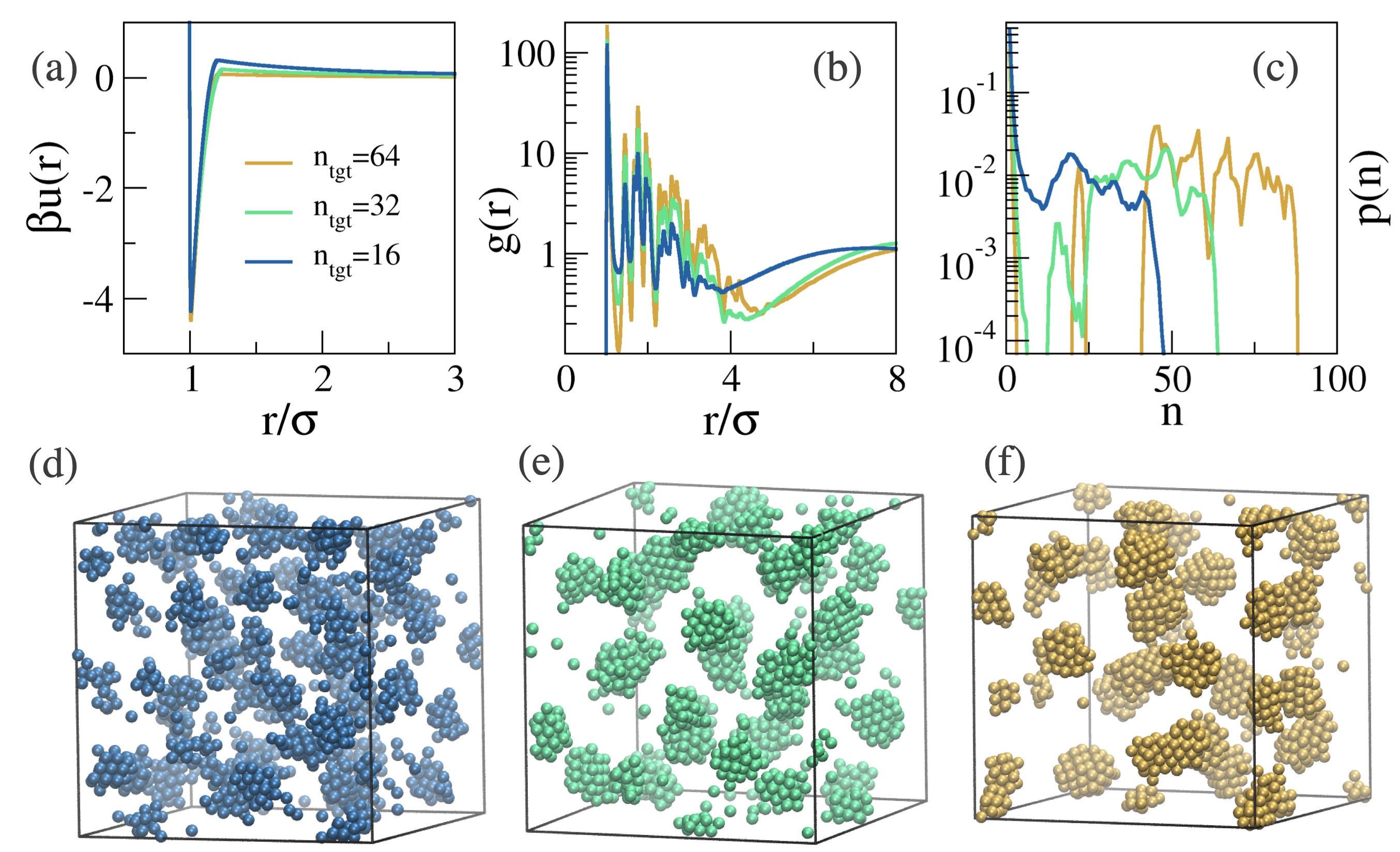}
  \caption{(a) Pair potentials, (b) radial distribution functions, and (c) cluster size distributions optimized for $n_{\text{tgt}}=16, 32, 64$. (d-f) From smallest to largest $n_{\text{tgt}}$, representative snapshots from the production simulations.}
  \label{fgr:size}
\end{figure}

Previous work has shown that inverse design is capable of optimizing for pairwise isotropic interaction potentials that prompt self-assembly of clusters with a prescribed size when the form of the potential is infinitely flexible.~\cite{IBI_clusters} To assess the size specificity possible with the AO+DLVO interaction, we ran three separate target simulations with $n_{\text{tgt}}=16, 32, 64$ and performed the RE minimization for each case. The optimized pair interactions, and corresponding radial distribution functions and cluster size distributions, are shown in Fig.~\ref{fgr:size}. Excluding the monomers and small ($n \le 7$) clusters, the average cluster sizes, $\langle n \rangle$, are 24.2, 41.5, and 61.0. While the correct qualitative trend is observed, the relative deviation between $\langle n \rangle$ and $n_{\text{tgt}}$ grows as the cluster size decreases. Consistent with the notion that cluster size is limited by building up sufficient repulsions to inhibit further particle addition, Fig.~\ref{fgr:size}a shows that the magnitude and range of the repulsive component is inversely related to $n_{\text{tgt}}$. In Fig.~\ref{fgr:size}b, it is clear that the rarefied zone related to intercluster spacing shifts to larger values of $r$ with increasing $n_{\text{tgt}}$, located at $\approx 3.6\sigma$, $4.0\sigma$, and $4.7\sigma$, respectively. These locations are larger than the corresponding values in the target simulations, in keeping with $\langle n \rangle$ being larger than $n_{\text{tgt}}$ in most cases. Moreover, the depletion zone is less pronounced in magnitude relative to the idealized target clusters. Taken together, we conclude that the above optimization scheme is capable of achieving qualitative control of the cluster size. While more exact matching with respect to cluster size might be possible for different target simulations, precisely how the target simulations should be adjusted to enhance the agreement is unclear. 

The cluster size distributions in Fig.~\ref{fgr:size}c show the qualitative trend of increasing $n_{\text{tgt}}$; the relatively discrete nature of the distribution arises because the clusters appear to be relatively static, consistent with prior work on crystalline clusters.~\cite{SALR4} While the clusters do appear to have some ability to re-arrange, this process seems to be relatively slow. Representative snapshots from simulations with the optimized potentials are shown in Fig.~\ref{fgr:size}d-f, where the intracrystalline nature of the clusters is visually evident. At the center of mass level though, the clusters are disordered and move fluidically in the simulations box.

\subsection{Effects of target simulation}

\begin{figure}[!htb]
  \includegraphics[width=16cm, keepaspectratio]{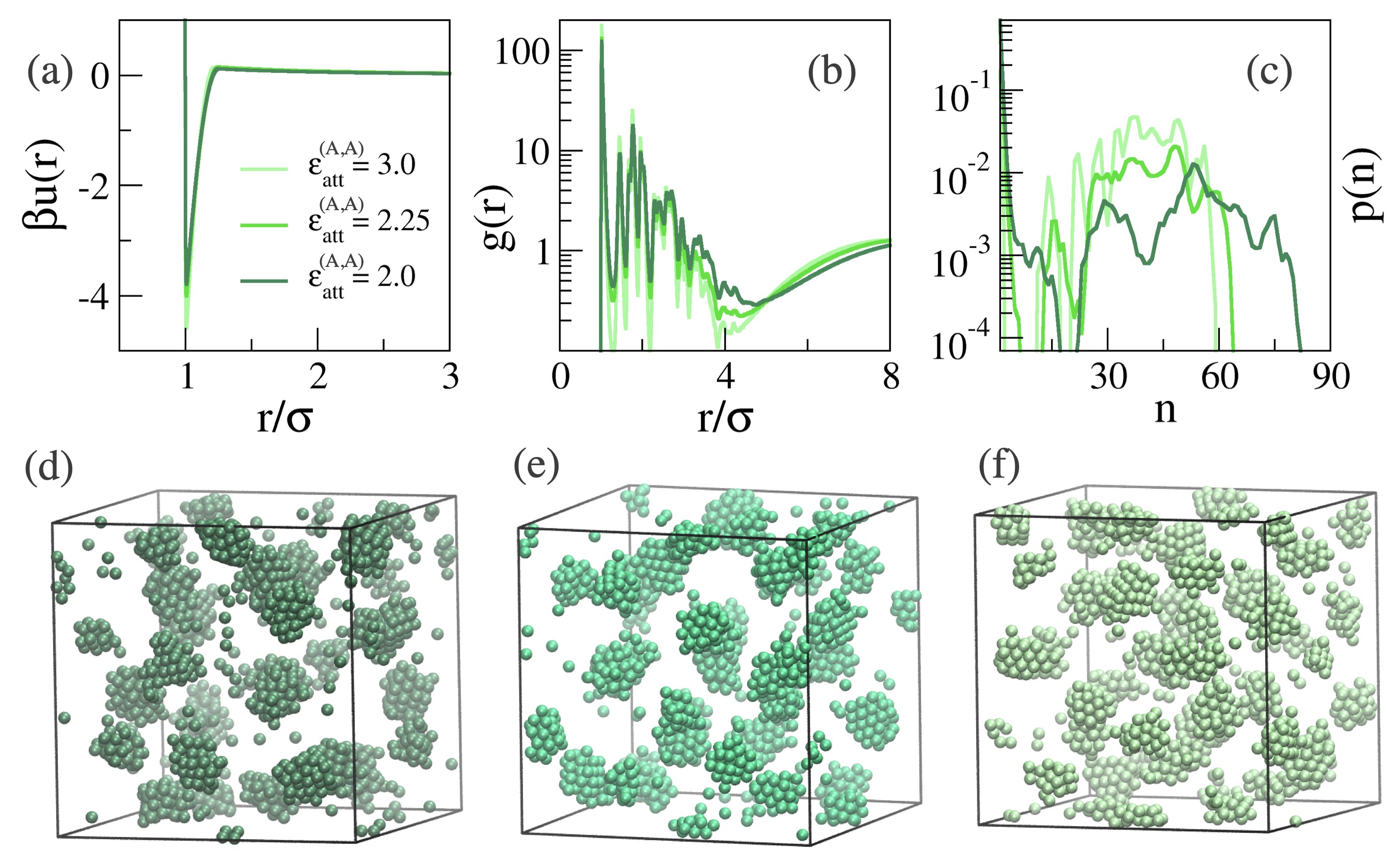}
  \caption{(a) Pair potentials, (b) radial distribution functions, and (c) cluster size distributions optimized for $\epsilon_{\text{att}}^{(A,A)}=2.0, 2.25, 3.0$. $\epsilon_{\text{att}}^{(A,A)}$ is directly related to the degree of intracluster crystallinity in the target simulation. (d-f) From from lowest to highest degree of crystallinity in the target simulations, representative snapshots from the production simulations.}
  \label{fgr:tgt_effects}
\end{figure}

For the target simulations used to generate the results shown in Fig.~\ref{fgr:size}, the parameters were selected such that, at the center-of-mass level, the clusters moved as a fluid; however, the particles inside a given cluster had some crystalline structure (see Fig.~\ref{fgr:tgt} as an example). This was done intentionally because the effective AO model is only valid when the range of the attraction is small, which favors crystallization over liquid condensation for purely attractive systems. In Fig.~\ref{fgr:tgt_effects}, we investigate whether the degree of intracluster crystallinity in the target simulation affects the outcome of the optimization for the case of $n_{\text{tgt}}=32$. By modulating the strength of attraction between the A particles in the target simulation, the degree of crystallinity in the target clusters is changed. The $n_{\text{tgt}}=32$ simulation in Fig.~\ref{fgr:size} was generated with $\epsilon_{\text{att}}^{(A,A)}=2.25$. For the more structured target, $\epsilon_{\text{att}}^{(A,A)}$ was set to 3.0; the less structured target, where the internal structure of the clusters was largely fluidic, had $\epsilon_{\text{att}}^{(A,A)}=2.0$. While the optimized potentials in Fig.~\ref{fgr:tgt_effects}a look relatively similar, greater crystallinity leads to a deeper attractive well coupled with a slightly larger repulsive barrier and vice versa. The differences in the degree of structuring are evident in $g(r)$ in Fig.~\ref{fgr:tgt_effects}b. As the structuring of the target increases, we see reduced population of monomer and other small clusters as well as increased monodispersity of the clusters, with the $\epsilon_{\text{att}}^{(A,A)}=2.0$ case being particularly polydisperse; see Fig.~\ref{fgr:tgt_effects}c. The two crystalline targets result in potentials that are somewhat size-specific; the average cluster sizes are 39.8 and 41.5 for $\epsilon_{\text{att}}^{(A,A)}=3.0, 2.25$, respectively. However, for the $\epsilon_{\text{att}}^{(A,A)}=2.0$ target simulation where the clusters are internally fluidic, the optimized clusters are significantly larger, with $\langle n \rangle=50.3$. This shift to larger clusters for $\epsilon_{\text{att}}^{(A,A)}=2.0$ is also apparent in the location of the rarefied zone in $g(r)$ shifting to larger $r$, as seen in Fig.~\ref{fgr:tgt_effects}b. 

From both the $g(r)$ in Fig.~\ref{fgr:tgt_effects}b and the snapshot shown in Fig.~\ref{fgr:tgt_effects}d, it is clear that clusters simulated with the potential optimized for the $\epsilon_{\text{att}}^{(A,A)}=2.0$ target are relatively crystalline. From comparing the target and optimized $g(r)$, it is clear that the optimized clusters are significantly more crystalline than in the target simulation (data not shown). We hypothesize that it is not possible to form internally fluidic clusters (at least when $n_{\text{tgt}}=32$) for a single-component simulation using an AO+DLVO potential where $q\le0.25$; therefore, it is not possible to match both the internal structure and the size of the clusters in the target ensemble simultaneously. The trade-off between these two competing factors may lead to the greater deviation of $\langle n \rangle$ from the target than when the target simulation possesses internally crystalline clusters. Taken together, these results emphasize that a realistic target simulation is necessary. Furthermore, as the optimized potential form becomes more restrictive, compatibility between the target and the model is expected to become even more crucial to meet design goals. While intuition is often a useful guide, selection of an optimal target in a systematic fashion for an arbitrary design goal remains an open challenge.    

The optimized parameters derived from all of the target simulations are given in Table~\ref{tbl:params}. The parameter $q$ is a size ratio that directly relates to experimental measurements. The parameter $\eta_p^r$ sets the chemical potential and therefore a grand canonical Monte Carlo simulation could be used to arrive at an actual packing fraction of polymer, $\eta_p$, that could be input to experiment. At the low colloidal packing fraction of 0.05 used here, a mean-field approximation, where the volume of available space that the polymers may occupy (if correlations between colloids are neglected) is filled in with a packing fraction of $\eta_p^r$, is probably quite reasonable and yields $\eta_p \approx 0.55-0.6$ for the above values, which is a physically reasonable range since the depletants in general can overlap with one another. To give a sense of how the optimized repulsive parameters might relate to experimental parameters, we assume a Bjerrum length $\lambda_B=0.7$nm, corresponding to water at room temperature, and we take $\sigma=50$nm. The screening lengths are rather large, yielding a low (6--10 $\mu$M) salt concentration of monovalent ions for the above values of $\lambda_B$ and $\sigma$; this value will further decrease with increasing $\sigma$. The values of $A$ for an equi-sized nanoparticle correspond to rather low effective charge density of $Z=3-6$, though this value with grow with $\sigma$.

\begin{table}[h]
\caption{\label{tbl:params}AO+DLVO parameters optimized for each target simulation (specified by $n_{\text{tgt}}$ and $\epsilon_{\text{att}}^{(A,A)}$)} 
\begin{center}
\lineup
\begin{tabular}{*{6}{l}}
\br                              
$n_{\text{tgt}}$&$\epsilon_{\text{att}}^{(A,A)}$&$\eta_p^r$&\m$q$&\m$A$&\m$z/\sigma$\cr 
\mr
16&4.00&0.629&0.213&0.574&3.005\cr
32&2.25&0.650&0.250&0.269&3.459\cr
64&2.25&0.620&0.218&0.115&3.134\cr
32&3.00&0.660&0.218&0.264&2.611\cr
32&2.00&0.612&0.250&0.214&3.343\cr
\br
\end{tabular}
\end{center}
\end{table}

\section{Conclusions} \label{sec:conclusions}
In this proceeding, we have demonstrated that the RE framework can be used to discover experimentally relevant parameters that lead to the self-assembly of clusters that are approximately of a specified size. The form of the experimentally relevant probabilistic model is relatively restrictive; as a result, we find that the target simulation must possess structural features that are consistent with the model constraints in order to achieve reasonable size-specificity. The approach outlined above can be further tailored to accommodate additional experimental constraints. For instance, if it is desirable for the particle charge to be a fixed quantity, then $\eta_p^r$, $q$, and $z$ may be optimized at a fixed value of $A$. Moreover, the value that any parameter is allowed to possess can also be further constrained, useful if a certain parameter range is known to be more compatible in some way for a given experimental system.

In future work, to better connect with experiment, it may well be more realistic to use a polydisperse mixture of particles to thwart crystallization within the clusters,~\cite{SALR4} as experimental realization of clustered systems are generally not crystalline. Similarly, more sophisticated models for the interactions are possible, including multi-body models that distinguish between surface particles and bulk particles.~\cite{renormalized_potential} Finally, the RE framework can be extended to the grand canonical ensemble to enable use of explicit models for depletion to study regimes where the effective AO model does not apply.

\ack
This work was supported by the Darleane Christian Hoffman Distinguished Postdoctoral Fellowship at Los Alamos National Laboratory.

\appendix
\section*{Appendix: Interactions for the Target Simulations}
\setcounter{section}{1}


The WCA interaction is~\cite{HansenMcDonald}

\begin{equation} \label{eqn:wca} 
\beta u_{\text{wca}}^{(i,j)}(r) \equiv H(2^{1/\alpha_{\text{wca}}^{(i,j)}} \sigma^{(i,j)}-r) \Bigg(4 \epsilon_{\text{wca}} \bigg[\bigg(\dfrac{\sigma^{(i,j)}}{r}\bigg)^{2\alpha_{wca}^{(i,j)}}-\bigg(\dfrac{\sigma^{(i,j)}}{r}\bigg)^{\alpha_{wca}^{(i,j)}}\bigg]+\epsilon_{\text{wca}}\Bigg)
\end{equation} 
and the form of the attractive wells is given by
\begin{equation} \label{eqn:att} 
\beta u_{\text{att}}^{(i,j)}(r) \equiv -\epsilon_{\text{att}}^{(i,j)} \text{exp} \bigg[ {-\Big(\frac{r-\sigma^{(i,j)}-\Delta^{(i,j)}}{\zeta^{(i,j)}}\Big)^{\alpha_{att}^{(i,j)}}} \bigg].
\end{equation}
where $\sigma^{(A,A)}=\sigma$, $\epsilon_{\text{wca}}=1$, $\epsilon_{\text{att}}^{(A,B)}=25$, $\zeta_{\text{att}}^{(A,A)}=\Delta_{\text{att}}^{(A,A)}=0.025$, $\alpha_{\text{att}}^{(A,A)}=6$, $\alpha_{\text{att}}^{(A,B)}=12$, $\alpha_{\text{wca}}^{(A,A)}=48$, and $\alpha_{\text{wca}}^{(A,B)}=24$. The remaining values that vary as a function of target simulation are given in Table A1, and any value not listed above or in the table should be taken to be zero. Though the exact values of the parameters are somewhat arbitrary, the values for $\sigma^{(B,B)}$ and $\zeta^{(A,B)}$ were inspired by Ref.~\citenum{IBI_clusters}.

\begin{table}[h]
\caption{\label{tabone}Target simulation parameters.} 
\begin{center}
\lineup
\begin{tabular}{*{4}{l}}
\br                              
$n_{\text{tgt}}$&$\epsilon_{\text{att}}^{(A,A)}$&\m$\sigma^{(B,B)}$&\m$\zeta^{(A,B)}$\cr 
\mr
16&4.00&$3.98\sigma$&$1.35\sigma$\cr
32&2.25&$5.60\sigma$&$1.85\sigma$\cr 
64&2.25&$6.70\sigma$&$2.30\sigma$\cr 
32&3.00&$5.60\sigma$&$1.85\sigma$\cr 
32&2.00&$5.60\sigma$&$1.85\sigma$\cr 
\br
\end{tabular}
\end{center}
\end{table}

\section*{References}

\bibliographystyle{iopart-num}
\providecommand{\newblock}{}

\end{document}